\documentstyle[12pt,epsfig]{article}

\begin{document}

\title{\vspace{1.5cm} Collision integral for multigluon production in a 
model for scalar quarks and gluons}

\author{D.~S. Isert and S.~P.~Klevansky\footnote{Current address: DB AG,
Kleyerstr.27, 60236 Frankfurt a.~M., Germany}\\
\small{Institut f\"ur Theoretische Physik,} \\
\small{Philosophenweg 19, D-69120 Heidelberg, Germany}\\
\small{E-mail: D.Isert@tphys.uni-heidelberg.de}}

\maketitle
\begin{abstract}
A model of scalar quarks and scalar gluons is used to derive 
transport equations for quarks and gluons. In particular, the collision
integral is studied. The self-energy diagrams are organized according to the
number of loops. 
A generalized Boltzmann equation is obtained, which involves 
at the level up to two loops all possible two $\to$ two parton
scattering processes and corrections to the process $q\bar q\to g$.
\end{abstract}

\section{Model of scalar quarks and gluons, and transport theory}

The model underpinning our calculations
successfully gives a $\ln s$ behavior in high energy $q q$
scattering~\cite{forsh}. Quarks and antiquarks are described by complex
scalar fields $\phi$, and gluons as the  scalar field $\chi$ coupled
through the Lagrangian
\begin{eqnarray}
{\cal L}&=& \partial^{\mu}{\phi}^{\dagger i,l}\partial_{\mu}\phi_{i,l} +
\frac{1}{2}\partial^{\mu}\chi_{a,r}\partial_{\mu}\chi^{a,r}
-\frac{m^2}{2}\chi_{a,r}\chi^{a,r} \nonumber \\
&& - g m{\phi}^{\dagger i,l}(T^a)^j_i(T^r)^m_l\phi_{j,m}\chi_{a,r} -
\frac{gm}{3!} f_{abc}f_{rst} \chi^{a,r}\chi^{b,s}\chi^{c,t}.
\label{e:lagrangian}
\end{eqnarray}
The quarks are massless, while the gluons are assigned  a mass $m$
in order to avoid infra-red divergences.
Since in QCD the quartic interaction
between gluons leads to terms which are sub-leading in ln $s$, it
is not included here.
The two labels on each of the fields refer to the direct product of two
color groups which is necessary to make the three gluon vertex
symmetric under the exchange of two bosonic gluons.

To describe non-equilibrium phenomena, we use the Green functions of the
Schwinger-Keldysh formalism~\cite{schwing}. The equations of motion for the
Wigner transforms of the Green functions, the so-called transport and 
constraint equations, are derived~\cite{ulli,ogu} and read
\begin{equation}
-2ip^{\mu}\partial_{X\mu} D^{-+}(X,p)=I_{\rm coll} + I_-^A + I_-^R
\quad\quad {\rm transport}
\label{e:transport}
\end{equation}
and
\begin{equation}
\left( \frac{1}{2} \partial_X^{\mu}\partial_{X\mu} 
- 2p^2 +2M^2\right) D^{-+}(X,p) = I_{\rm coll} + I_+^A + I_+^R,
\quad\quad
{\rm constraint},
\label{e:constraint}
\end{equation}
where $D^{-+}$ is a generic Green function, $D^{-+}=S^{-+}$ for
quarks and $D^{-+}=G^{-+}$ for gluons. $M$ is the appropriate mass, and
$I_{\rm coll}$ is the collision term,
\begin{equation}
I_{\rm coll} = \Pi^{-+}(X,p)\hat{\Lambda}D^{+-}(X,p)
                 -\Pi^{+-}(X,p)\hat{\Lambda}D^{-+}(X,p)
             = I_{\rm coll}^{\rm gain} - I_{\rm coll}^{\rm loss}.
\label{e:collision}
\end{equation}
$I_{\mp}^R$ and $I_{\mp}^A$ are  terms containing retarded and
advanced components:
\begin{eqnarray}
I_{\mp}^R &=& -\Pi^{-+}(X,p)\hat{\Lambda}D^{R}(X,p)
            \pm    D^{R}(X,p)\hat{\Lambda}\Pi^{-+}(X,p)\\
\label{e:rim}
I_{\mp}^A &=& \Pi^{A}(X,p)\hat{\Lambda}D^{-+}(X,p)
            \mp    D^{-+}(X,p)\hat{\Lambda}\Pi^{A}(X,p).
\label{e:pim}
\end{eqnarray}
In Eqs.(\ref{e:collision}) to (\ref{e:pim}), $\Pi$ is a generic self-energy
($\Sigma_q$ for quarks and $\Sigma_g$ for gluons) and
the operator $\hat \Lambda$ is given by
\begin{equation}
\hat{\Lambda}:={\rm exp}\left\{ \frac{-i}{2}\left(\overleftarrow{\partial}_X
 \overrightarrow{\partial}_p
-\overleftarrow{\partial}_p\overrightarrow{\partial}_X\right) \right\}.
\label{lambda1}
\end{equation}
For the Green functions, we use the quasiparticle approximation:
\begin{eqnarray}
iD^{-+}(X,p)&=&\frac{\pi}{E_p}\{\delta(E_p-p^0)f_a(X,p)+\delta(E_p+p^0)
\bar{f}_{\bar a} (X,-p)\}\label{d-+}\\
iD^{\mp \mp}(X,p)&=&\frac{\pm i}{p^2-M^2\pm i\epsilon}
+\Theta(-p^0)iD^{+-}(X,p)+ \Theta(p^0)iD^{-+}(X,p)\label{d--},
\end{eqnarray}
$D^{+-}$ is obtained from the expression for $D^{-+}$ by replacing $f$ with
$\bar f$ and vice versa. $f_a$ is the quark ($a=q$) or the gluon ($a=g$)

distribution function, $\bar f_a$ is an abbreviation for $1+f_a$.
Then an evolution equation for the parton distribution function is obtained
by integrating Eq.(\ref{e:transport}) over an interval $\Delta^+$ that 
contains the energy $E_p$. To lowest order in an expansion that sets 
$\hat \Lambda = 1$, the integral for the collision term is performed in the 
next section.

\section{The collision integral}

For the collision integral, one has
\begin{eqnarray}
J_{\rm coll}
&=& \int_{\Delta^+}\! dp_0\,\Pi^{-+}(X,p)\,D^{+-}(X,p)
   -\int_{\Delta^+}\! dp_0\,\Pi^{+-}(X,p)\,D^{-+}(X,p)\nonumber\\
&=& -i
\frac{\pi}{E_p}\,\Pi^{-+}(X,p_0\!=\!E_p,\vec{p})\,\bar{f}_a(X,\vec{p})
    +i \frac{\pi}{E_p}\,\Pi^{+-}(X,p_0\!=\!E_p,\vec{p})\,f_a(X,\vec{p}),
\nonumber\\
\label{e:collint}
\end{eqnarray}
{\it i.e.} the off-diagonal quasiparticle self-energies are required to be
calculated on shell. 

In the following, only the results for the quarks are presented, 
the generalization to the gluons is then obvious.
The self-energies are organized now according to their number of loops.

\subsection{Hartree and Fock self-energies}
The Hartree self-energies (self-energy with a quark or a gluon loop) 
vanish due to their color factors as it is the case for real QCD.

The Fock self-energy gives a contribution to the loss term of the collision
integral in Eq.(\ref{e:collint}) given by~\cite{isert}
\begin{eqnarray}
J_{\rm coll}^{\rm loss}&=& -\frac{\pi}{E_p} \int\frac{d^3p_1}{(2\pi)^3 2E_1}
\frac {d^3p_2}{(2\pi)^3
2E_2} (2\pi)^4 \delta^{(4)}(p+p_1-p_2)\nonumber \\
&& \times |{\cal M}_{q\bar q\to g}|^2 f_q(X,\vec p)f_{\bar q}(X,\vec p_1)
\bar f_g(X,\vec p_2).
\label{e:lossint}
\end{eqnarray}
Here, ${\cal M}_{q\bar q\to g}$ means the scattering amplitude of the
process $q\bar q\to g$ of order $gm$ which is pictured in Fig.1 a).
Processes as $qg\to q$, $q\to qg$ or $q\bar qg\to${\O} are kinematically
forbidden and therefore do not occur in this collision integral.
The gain term can be obtained by exchanging $f$ with $\bar f$ in
Eq.(\ref{e:lossint}).
\begin{figure}[h]
\centerline{\scalebox{0.8}{\includegraphics{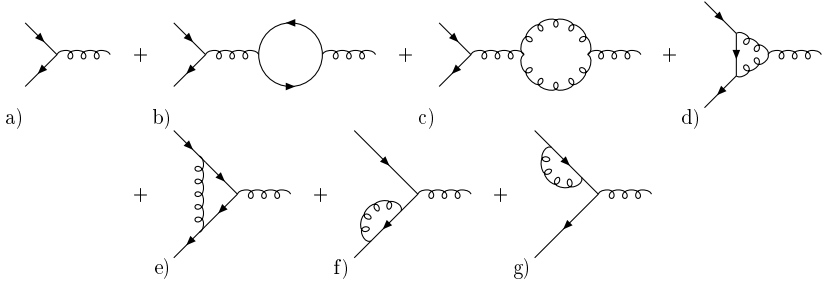}}}
\caption{The process $q\bar q\to g$ up to order $g^3m^3$}
\end{figure}

\subsection{Two loop self-energies}
The two loop self-energy graphs are pictured in Fig.2. According to their
topology, they are called rainbow (R), ladder (L), cloud (C), exchange (E)
and quark-loop (QL) graphs. 

\begin{figure}[h]
\centerline{\scalebox{0.8}{\includegraphics{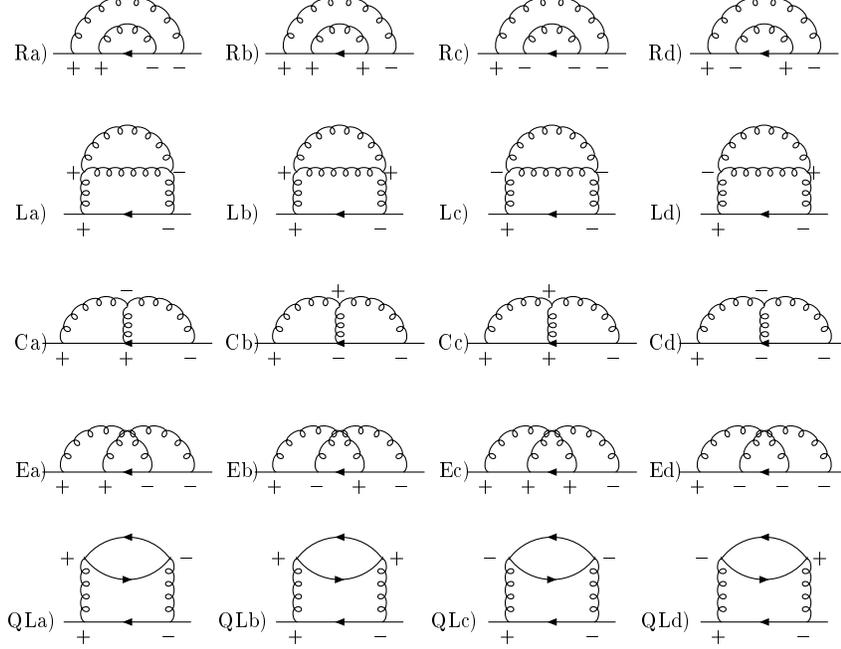}}}
\caption{Two loop self-energy diagrams}
\end{figure}

Let us first consider the contribution of the seven diagrams  Ra), La), Ca), 
Cb), Ea), Eb) and QLa) to the collision integral. 
The first five of these diagrams lead to the (lowest order) 
scattering amplitudes of the processes $q\bar q \to gg$ and $qg \to qg$, 
while the latter two lead to the (lowest order) scattering amplitudes of 
the processes $qq\to qq$ and 
$q\bar q\to q\bar q$~\cite{isert}. Thus, to obtain {\it all} possible 
$2\to2$ processes it is necessary to consider all types of diagrams.
But an analysis of the color factors shows that the quark-loop diagrams are
subleading as it is the case in real QCD. 
Therefore, in an additional expansion in the inverse number of
colors, $1/N_c$, one can neglect the processes $qq\to qq$ and 
$q\bar q\to q\bar q$. That means that gluon production is favored.

The three diagrams Rd), Ld) and QLd) vanish due to the momentum structure of 
their propagators~\cite{isert2}.
 
To see the purpose of the remaining ten diagrams~\cite{isert2}, i.e.
Rb), Rc), Lb), Lc), Cc), Cd), Ec), Ed), QLb) and QLc), we 
investigate first the process $q\bar q\to g$
which is shown up to order $g^3m^3$ in Fig.1. 
$|{\cal M}_{q\bar q\to g}|^2$ was given in lowest order by the Fock term. To
construct it up to order $g^4m^4$, one has to take the hermitian conjugate
of the scattering amplitude of Fig.1a) and multiply it with 
each of the scattering amplitudes of Fig.1b)-g).
It is also necessary to take 
the hermitian conjugate of each of these products, too. Using
a symbolical notation, we write these products as $a^{\dagger}b,
ab^{\dagger}, a^{\dagger}c, ...$. Then a detailed analysis shows that each 
of these products, with the exception of $a^{\dagger}g$ and $ag^{\dagger}$, 
is provided by one of the above mentioned ten self-energy diagrams. 
The diagram g) of Fig.1 does not enter into the collision integral, as it 
is a renormalization diagram for the incoming quark, for which the momentum 
$p$ is fixed externally.  

So far, we have constructed the collision integral out of self-energy
diagrams with graphs up to two loops. Let us now return to the transport 
equation
(\ref{e:transport}). If we integrate the left hand side over the interval
$\Delta^+$ and neglect the contributions of $I_A$ and $I_R$ on the right
hand side, then we get for the transport equation up to order $g^4m^4$
\begin{eqnarray}
&&2p^{\mu}\partial_{X\mu}f_q(X,\vec{p})=
\int\frac{d^3p_1}{(2\pi)^3 2E_1}\frac {d^3p_2}{(2\pi)^3 2E_2} 
(2\pi)^4 \delta^{(4)}(p+p_1-p_2)\nonumber \\
&& \times |{\cal M}_{q\bar q\to g}|^2 \left\{\bar f_q(X,\vec p)
\bar f_{\bar q}(X,\vec p_1)f_g(X,\vec p_2) -
f_q(X,\vec p)f_{\bar q}(X,\vec p_1)\bar f_g(X,\vec p_2)\right\}\nonumber\\
&& + \int\frac{d^3p_1}{(2\pi)^3 2E_1}\frac {d^3p_2}{(2\pi)^3 2E_2} 
\frac{d^3p_3}{(2\pi)^3 2E_3}
(2\pi)^4 \delta^{(4)}(p+p_1-p_2-p_3)\nonumber \\
&&\times \left[\frac{1}{2} |{\cal M}_{q\bar q\to gg}|^2 \left\{\bar
f_q(X,\vec p)\bar f_{\bar q}(X,\vec p_1)f_g(X,\vec p_2)f_g(X,\vec p_3)
\right.\right.\nonumber\\
&&\left.\qquad\qquad\qquad\; -f_q(X,\vec p)f_{\bar q}(X,\vec p_1) 
\bar f_g(X,\vec p_2) \bar f_g(X,\vec p_3)\right\}\nonumber\\
&&\;\;\;+ |{\cal M}_{qg\to qg}|^2 \left\{\bar f_q(X,\vec p)
\bar f_g(X,\vec p_1) f_q(X,\vec p_2) f_g(X,\vec p_3) \right.\nonumber\\
&&\left.\left. \qquad\qquad\qquad- 
f_q(X,\vec p) f_g(X,\vec p_1) \bar f_q(X,\vec p_2)
\bar f_g(X,\vec p_3)\right\}\right].
\end{eqnarray}

The generalization to higher orders is then obvious. 
Including, e.g., the three loop self-energy into the collision term
leads to cross sections of all kinematically allowed 
processes with two (three) partons in the initial state and three (two)
partons in the final state. Furthermore, they give all additional
corrections of order $g^6m^6$ to the process $q\bar q\to g$, 
e.g.~the product of scattering amplitudes of the diagrams b)-f) in Fig.2, 
and also corrections to the scattering processes of two partons into two 
partons.

\section{Closing comments}
One may speculate that pinch singularities could enter into the collision
integral up to the two loop level. In fact this is not so, and has been
explicitly demonstrated~\cite{isert2}.

\end{document}